\documentclass[12pt]{article}
\usepackage[margin=1in]{geometry}
\usepackage{amsmath}
\usepackage{mathtools}
\usepackage{graphicx}
\usepackage{amsfonts}
\usepackage{amssymb}
\usepackage{enumitem}
\usepackage{nicefrac}
\usepackage{setspace}
\usepackage{titlesec}
\usepackage{textcomp}
\usepackage{natbib}
\usepackage{multirow}
\usepackage{caption}
\usepackage{authblk}
\usepackage[mathlines]{lineno}
\usepackage{soul}
\usepackage{xcolor}
\usepackage{url}
\usepackage[symbol]{footmisc}
\usepackage{helvet}
\usepackage{lineno}
\usepackage{hyperref}
\usepackage{verbatim}

\newcommand{\ecomem}{\texttt{EcoMem}}
\newcommand{\by}{\mathbf{y}}
\newcommand{\bx}{\mathbf{x}}

\newcommand{\bX}{\mathbf{X}}

\newcommand{\wj}{\mathbf{w}_{j}}
\newcommand{\bfj}{\mathbf{H}_{j}}
\newcommand{\bfcoefj}{\boldsymbol{\eta}_{j}}
\newcommand{\one}{\mathbf{1}}
\newcommand{\regj}{\tau_{j}^{2}}
\newcommand{\pmatj}{\mathbf{S}_{j}}

\title{\ecomem{}: An R package for quantifying ecological memory}

\author[1,2]{Malcolm Itter\footnote{Corresponding author. Email: malcolm.itter@helsinki.fi}}
\author[1,2]{Jarno Vanhatalo}
\author[3,4]{Andrew Finley}
\affil[1]{Research Centre for Ecological Change, Organismal and Evolutionary Biology Research Program, \protect\\ University of Helsinki, Helsinki, Finland}
\affil[2]{Department of Mathematics and Statistics, University of Helsinki, Helsinki, Finland}
\affil[3]{Department of Forestry, Michigan State University, East Lansing, Michigan, USA}
\affil[4]{Department of Geography, Michigan State University, East Lansing, Michigan, USA}

\date{}

\begin{document}

\maketitle

\begin{abstract}
\noindent{Ecological processes may exhibit memory to past disturbances affecting the resilience of ecosystems to future disturbance. Understanding the role of ecological memory in shaping ecosystem responses to disturbance under global change is a critical step toward developing effective adaptive management strategies to maintain ecosystem function and biodiversity. We developed \ecomem{}, an \texttt{R} package for quantifying ecological memory functions using common environmental time series data (continuous, count, proportional) applying a Bayesian hierarchical framework. The package estimates memory functions for continuous and binary (e.g., disturbance chronology) variables making no \textit{a priori} assumption on the form of the functions. \ecomem{} allows users to quantify ecological memory for a wide range of ecosystem processes and responses. The utility of the package to advance understanding of the memory of ecosystems to environmental drivers is demonstrated using a simulated dataset and a case study assessing the memory of boreal tree growth to insect defoliation.}
\end{abstract}

\section{Introduction}\label{sec:intro}
Ecological processes may exhibit ``memory'' to past conditions. That is, the current function of an ecosystem may be affected by exogenous (e.g., weather, disturbance, management) and endogenous (e.g., successional stages, community composition, functional diversity) factors over a range of past time points. Ecological memory is often used in the context of ecological responses to disturbance and the formation of resilient ecosystems \citep{Gunderson2000,Johnstone2016}. In this context, ecological memory is defined as the degree to which an ecosystem's response to a current or future disturbance is shaped by its responses to past disturbances \citep{Padisak1992,Peterson2002}. This includes biological legacies such as changes in the structure, function, and diversity of a system following disturbance \citep{Johnstone2016}. It also includes persistent responses to a disturbance, which may limit the capacity of the system to respond to future disturbance events \citep{Anderegg2015}. Ecological memory has been demonstrated in forest systems \citep{Anderegg2015}, freshwater phytoplankton \citep{Padisak1992}, coral reefs \citep{Nystrom2001}, and grasses \citep{Walter2011}.

Resilient ecosystems are well adapted to regional disturbance regimes \citep{Johnstone2016}. As disturbance events become more frequent and/or severe under global change, systems may accumulate stress over time such that future disturbance has an unexpectedly profound effect on ecosystem function. By accounting for the legacy effects of disturbance, ecological memory provides a mechanism to quantify the accumulation of stress within an ecosystem over repeated disturbance events. Identifying ecosystem attributes limiting the effects of disturbance over time as reflected by ecological memory is a critical step in the development of adaptive management strategies aimed at promoting resilient ecosystems under novel conditions \citep{Folke2004}.

\citet{Ogle2015} were the first to provide an integrated approach to quantify ecological memory. They define memory as comprising three components: i) the length of an ecosystem's response to previous conditions; ii) the relative importance of conditions at specific times in the past; and, iii) the strength of an ecosystem's response to temporally-integrated conditions. The Bayesian hierarchical model developed by \citet{Ogle2015} provides a flexible framework to quantify each of the components of ecological memory.

We extend the Bayesian hierarchical model developed by \citet{Ogle2015} and integrate the model into a new \texttt{R} package (\ecomem{}) for application to a wide-range of ecological time series data. Extensions to the model framework presented in \citet{Ogle2015} include developing an efficient and flexible spline-based approach to quantify ecological memory, and allowing for estimation of memory to both continuous covariates and binary event data (e.g., a disturbance chronology).

\section{Methods}\label{sec:methods}

\subsection{Setting}\label{subsec:setting}
Ecological memory is quantified through the estimation of latent weights reflecting the relative importance of past conditions on current ecosystem function \citep{Ogle2015}. These weights are used to construct temporally-averaged covariates to model ecological processes. Suppose we are interested in an ecological process $Y$ observed over a range of time points: $\by \equiv (y_{1},y_{2},\ldots,y_{T})^{\intercal}$. Coincidentally, we observe values for a set of $p$ covariates believed to impact the process of interest represented by a $T\times{p}$ matrix $\bX$ where the $t$th row is given by $\bx_{t}^{\intercal} \equiv (x_{t,1},x_{t,2},\ldots,x_{t,p})$ for $t=1,\ldots,T$. A common approach is to construct a model to estimate the process as a function of concurrent conditions, $g(\text{E}(Y_{t}\lvert{\bX})) = f(\bx_{t};\boldsymbol{\theta})$, where $g(\cdot)$ is a link function (e.g., identity, logit), $\text{E}(\cdot)$ indicates the expected value, and $\boldsymbol{\theta}$ is a set of potentially unknown model parameters. In the case where the ecological process is thought to depend on conditions over a series of past time points, $g(\text{E}(Y_{t}\lvert\bX))$ can be estimated as a function of lagged covariate values. For example in a regression model,
\begin{linenomath*}
\begin{equation}\label{eqtn:dist_lag}
    g(\text{E}(Y_{t}\lvert\bX)) = \mu + \sum_{\ell=0}^{L}\bx_{t-\ell}^{\intercal}\boldsymbol{\alpha}_{\ell}
\end{equation}
\end{linenomath*}
where $\mu$ is an intercept, $\ell$ indicates the time lag (up to a maximum $L$), and $\boldsymbol{\alpha}_{\ell}$ is a $p$-dimensional vector of regression coefficients corresponding to the $\ell$th lag. We assume that all covariates have the same maximum lag in Eq. (\ref{eqtn:dist_lag}) for simplicity, although this need not be the case. Eq. (\ref{eqtn:dist_lag}) defines a distributed lag model \citep{Zanobetti2000,Heaton2012}. When $L$ is large or the lagged covariate values ($\bx_{j,t},\bx_{j,t-1},\ldots,\bx_{j,t-L}$) are correlated, $\boldsymbol{\alpha}_{j} \equiv (\alpha_{j,0},\alpha_{j,1},\ldots,\alpha_{j,L})^{\intercal}$ is modeled jointly taking into account the covariance among coefficients.

An alternative approach in such settings first presented by \citet{Ogle2015}, is to filter covariate observations $\bx_{j}$, equivalent to the $j$th column of $\bX$ ($j=1,\ldots,p$), based on weights representing the relative importance of past conditions on the current process,
\begin{linenomath*}
\begin{equation}\label{eqtn:tildeX}
    \tilde{x}_{j,t} = \sum_{\ell=0}^{L}w_{j,\ell}x_{j,t-\ell}
\end{equation}
\end{linenomath*}
where $\tilde{x}_{j,t}$ is the filtered value of $\bx_{j}$ at time $t$ and $w_{j,\ell}$ is the weight corresponding to the $(t-\ell)$th lag of $\bx_{j}$. The temporally-filtered covariates are then be applied to model the ecological process at time $t$,
\begin{equation}\label{eqtn:wtd_reg}
    g(\text{E}(Y_{t}\lvert\bX)) = \mu + \tilde{\bx}_{t}^{\intercal}\boldsymbol{\beta}
\end{equation}
where $\tilde{\bx}_{t}$ and $\boldsymbol{\beta}$ are $p$-dimensional vectors of filtered covariate values and regression coeffecients, respectively. Eq. (\ref{eqtn:tildeX}) represents the construction of a temporally-filtered covariate in discrete time, but can be extended to continuous time. Constraints are placed on the weights in Eq. (\ref{eqtn:tildeX}) to ensure their idenfiability: (i) $w_{j,\ell} \in (0,1)$ for $\ell=0,1,\ldots,L$; (ii) $\sum_{\ell=0}^{L}w_{j,\ell}=1$. Note that without these constraints, Eq. (\ref{eqtn:wtd_reg}) is identical to the distributed lag model with $\boldsymbol{\alpha}_{j}=\beta_{j}\mathbf{w}_{j}$ given $\mathbf{w}_{j} \equiv (w_{j,0},w_{j,1},\ldots,w_{j,L})^{\intercal}$.

The set of weights for a given covariate ($\mathbf{w}_{j}$) defines an ``ecological memory function'' with several interpretations. The number of lags with weights above a specified lower bound (e.g., 0.01) indicates the length of a process' memory to the covariate. The magnitude of weights indicate the relative importance of covariate values at specific lags. The temporally-filtered covariates are similar in construction and interpretation to spatially-averaged covariates for use in regression models \citep{Heaton2011}, and reflect the accumulation of covariate values over the period $[0,L]$. The latent weights ($\mathbf{w}_{j}$) can be difficult to estimate given they are not placed directly on the response and may have complex correlation structure (see following section).

\subsection{Model framework}\label{subsec:model}
The initial approach to quantify ecological memory functions jointly estimates weights and fits a linear regression model using temporally-filtered covariates within a Bayesian hierarchical framework \citep{Ogle2015}. Weights are assigned a non-informative Dirichlet prior: $\wj \sim \mathcal{D}(\one)$. In cases where $L$ is large, it may be difficult to identify weights for all lags. We can improve weight estimation by reducing the dimension of the parameter space and imposing greater structure on the weights, which are likely to exhibit high temporal autocorrelation. Gaussian processes and penalized splines have been successfully applied to estimate coefficients in distributed lag models \citep{Zanobetti2000,Heaton2012}, and are natural candidates to estimate weights ($\wj$) given their close connection to lagged coefficients ($\boldsymbol{\alpha}_{j}$).

We apply a Bayesian hierarchical model utilizing penalized regression splines to estimate ecological memory functions. Specifically, weights are estimated as,
\begin{linenomath*}
\begin{equation}\label{eqtn:wt_est}
    \wj = \frac{\text{e}^{\bfj\bfcoefj}}{\one^{\intercal}\text{e}^{\bfj\bfcoefj}}
\end{equation}
\end{linenomath*}
where $\bfj$ is an $(L+1)\times{k}$ matrix containing $k$ spline basis function evaluations for each lag, $\bfcoefj$ is a $k$-dimensional vector of basis function coefficients, $\one$ is an $(L+1)$-dimensional vector of ones, and $\text{e}^{(\cdot)}$ defines a point-wise operation. Modeling weights on the log scale and the sum in the denominator term in Eq. (\ref{eqtn:wt_est}) ensure the identifiability constraints are met.

Consistent with penalized spline models, knots are placed within $[0,L]$ to model $\wj$ \citep{Wood2002}. Regularization is used to avoid overfitting and ensure the identifiability of the spline basis function coefficients ($\bfcoefj$). We define a weakly informative prior for the basis function coefficients, $\bfcoefj \sim \text{N}(\mathbf{0},\regj\pmatj^{-})$, where $\regj$ is a scalar variance parameter, $\pmatj$ is a $k\times{k}$ penalty matrix based on basis functions and knot locations, and $(^{-})$ indicates the generalized inverse as $\pmatj$ may not be full rank. The variance ($\regj$) serves as the regulator controlling the relative smoothness of the weights $\wj$. The weakly informative prior for $\bfcoefj$ is a form of shrinkage prior---when $\regj$ is small, the weight function is smooth.

Identifying optimal regulator values ($\regj$) is crucial for estimating meaningful weight functions ($\wj$). Prior distributions for regulator parameters in Bayesian penalized spline models are an active area of research. Recent work has proposed a set of priors for $\regj$ that incorporate a penalty for model complexity \citep{Ventrucci2016,Simpson2017}. In the context of ecological memory functions, we have found a folded $t$ distribution assigned to the square root of the regulator ($\tau_{j}$) allows for sufficiently-flexible basis function coefficients while controlling against overfitting similar to a penalized complexity prior.

We use Markov chain Monte Carlo (MCMC) to sample from the joint posterior distribution for the ecological memory model after specifying prior distributions for remaining model parameters \citep{RobertCasella2004}. Details on the priors used for all model parameters and the MCMC procedure (including efficient sampling considerations) are provided in Appendix A.

\section{Results}\label{sec:application}
The \ecomem{} package allows users to fit the model defined in Section \ref{sec:methods}. The development version of the package is available on GitHub and can be installed using \texttt{devtools} \citep{devtools}. A list of the core functions available within \ecomem{} is provided in Table \ref{tab:fn_list}. The current ecological memory model framework supports continuous, count, and proportional data utilizing Gaussian, Poisson, and binomial likelihoods, respectively. In all cases, the mean of the data is estimated according to Eq. (\ref{eqtn:wtd_reg}).

\subsection{Simulated example}
We demonstrate the steps necessary to quantify ecological memory using the \ecomem{} package through its application to a simulated dataset. The \texttt{mem.dat} dataset is included as part of the \ecomem{} package and contains Poisson count data, $\texttt{y} \sim \text{Poisson}(\lambda)$, that have been generated applying the ecological memory model using a log link function of two memory covariates (\texttt{v1},\texttt{v2}) and an auxiliary continuous covariate (\texttt{v3}), defined in \texttt{R} syntax,
\[\text{log}\;\lambda = \mu + \beta_{1}(\texttt{v1}) + 
\beta_{2}(\texttt{v2}) + \beta_{3}(\texttt{v3}) + \beta_{4}(\texttt{v1:v2}) + 
\beta_{5}(\texttt{v2:v3}),\]
where $\mu$ is an intercept term and the $\beta$'s are regression coefficients as defined in Eq. (\ref{eqtn:wtd_reg}). Code to generate the simulated dataset is provided in Appendix B. The \texttt{v2} variable is continuous while \texttt{v1} is binary indicating the occurrence of discrete events ($0=\text{no event}$, $1=\text{event}$).

\begin{table}
\centering
\caption{List of core functions available within the \ecomem{} package.}
\begin{tabular}{ll}
Function & Description\\\hline
\texttt{ecomem} & Quantifies ecological memory within a linear model framework\\
\multirow{2}{*}{\texttt{ecomemGLM}} & Quantifies ecological memory within a generalized linear\\ & model framework\\
\texttt{ecomemMCMC} & Implements MCMC inference for \texttt{ecomem} model\\
\texttt{ecomemGLMMCMC}& Implements MCMC inference for \texttt{ecomemGLM} model\\
\texttt{mem2mcmc} & Converts \texttt{ecomem} model object to \texttt{mcmc} object for use with \texttt{coda}\\
\texttt{memsum} & Summarizes marginal posterior distributions of \texttt{ecomem} parameters\\
\texttt{plotmem} & Plots estimated ecological memory functions\\\hline
\end{tabular}
\label{tab:fn_list}
\end{table}

\subsubsection{Fitting the model}
We apply \texttt{ecomemGLM()} to fit the ecological memory model to the Poisson count data. The \texttt{ecomemGLM()} call requires users to specify a linear model formula, the likelihood function for the data (Poisson or binomial), a model data frame including the response, explanatory variables, and all auxiliary variables, a subset of covariates for which memory functions should be estimated, the maximum lag for each memory covariate, as well as time and, if applicable, group identifiers (the group identifier is used if time series data exist for separate groups). The model is fit using response data for which explanatory variable observations exist for at least \texttt{max(L)} previous time points.

\begin{verbatim}
# Fit ecological memory model
mod = ecomemGLM(y ~ v1*v2 + v2*v3, family = "poisson",
                data = mem.dat, mem.vars = c("v1","v2"), 
                L = c(10,6), timeID = "time", groupID = "group")
\end{verbatim}

\subsubsection{Model outputs}
\begin{figure}
\centering
\includegraphics[width=\textwidth]{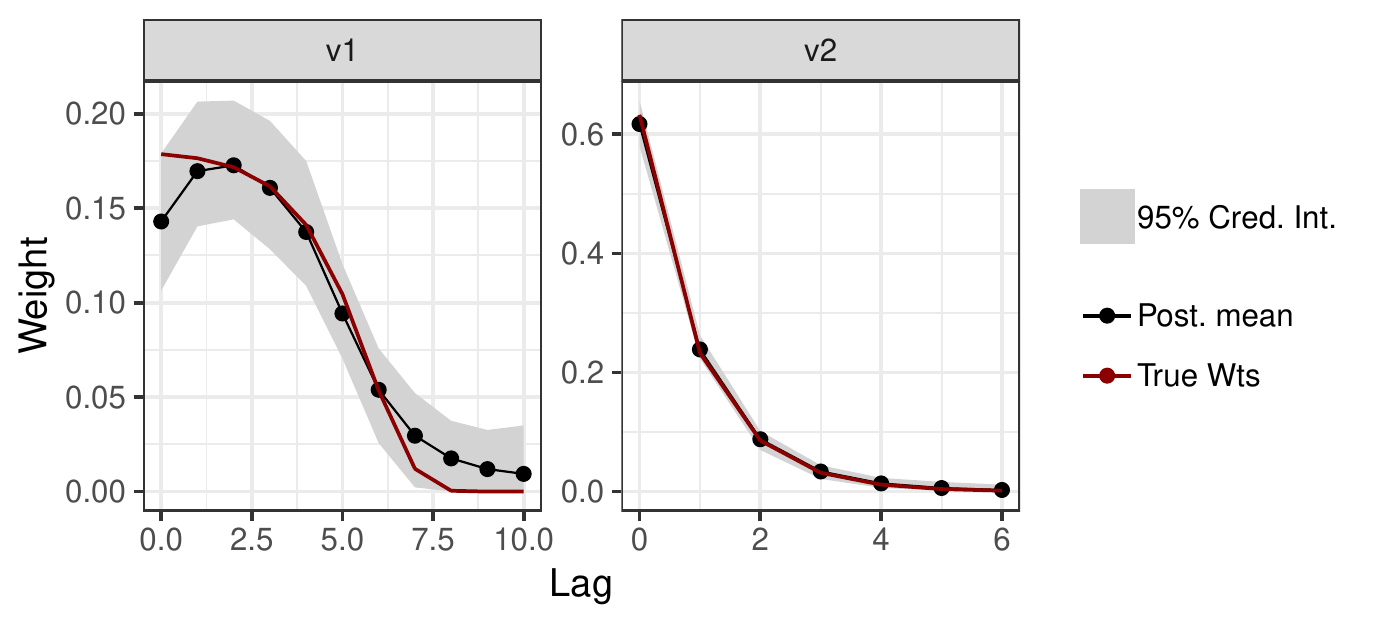}
\caption{Ecological memory functions for \texttt{v1} and \texttt{v2} covariates generated using the \texttt{plotmem()} function modified to display the data-generating weight functions for reference.}
\label{fig:plotmem}
\end{figure}

The \texttt{ecomemGLM()} and \texttt{ecomem()} functions return a list of class \texttt{ecomem} including posterior samples for each MCMC chain and the data used to fit the model (continuous covariates are standardized to have mean zero and variance one prior to model fitting). The posterior samples can be converted into an \texttt{mcmc} object using the \texttt{mem2mcmc()} function allowing users to apply the \texttt{coda} package to assess convergence \citep{Coda2006}. The marginal posterior distribution for each ecological memory model parameter can be summarized using the \texttt{memsum()} function (see Appendix B). Finally, ecological memory functions can be plotted for each memory covariate using the \texttt{plotmem()} function (Fig. \ref{fig:plotmem}).

\begin{verbatim}
# Plot memory functions
p = plotmem(mod, cred.int = 0.99)

p + geom_line(aes(y = wt, color = "True Wts"), size = 0.8,
              data = trueWts)  +
              scale_color_manual(values = c("black","darkred"))
\end{verbatim}

\subsection{Case study - boreal tree growth}
The utility of the \ecomem{} package for ecological inference is demonstrated through a case study assessing the memory of boreal tree growth to insect defoliation events. Recent work has shown that boreal tree growth exhibits negative responses to insect defoliation for several years following a moderate-to-severe defoliation event \citep{Itter2018}. We apply the \ecomem{} package to annual tree growth and insect defoliation survey data for 34 sites across Alberta, Canada to assess the ecological memory of tree growth to defoliation \citep[see][for detailed description of dataset]{Itter2018}. We model mean annual basal area increment of trembling aspen (\textit{Populus tremuloides} Michx.) as a function of forest tent caterpillar (\textit{Malascosoma disstria} Hub.) defoliation events and mean tree age allowing for memory to defoliation (additional details in Appendix B).

\begin{verbatim}
# Fit ecological memory model
mod = ecomem(gr ~ age + ftc, data = boreal.dat, mem.vars = "ftc",
             L = 12, timeID = "Year", groupID = "Stand")

plotmem(mod)
\end{verbatim}

\begin{figure}
\centering
\includegraphics[width=0.6\textwidth]{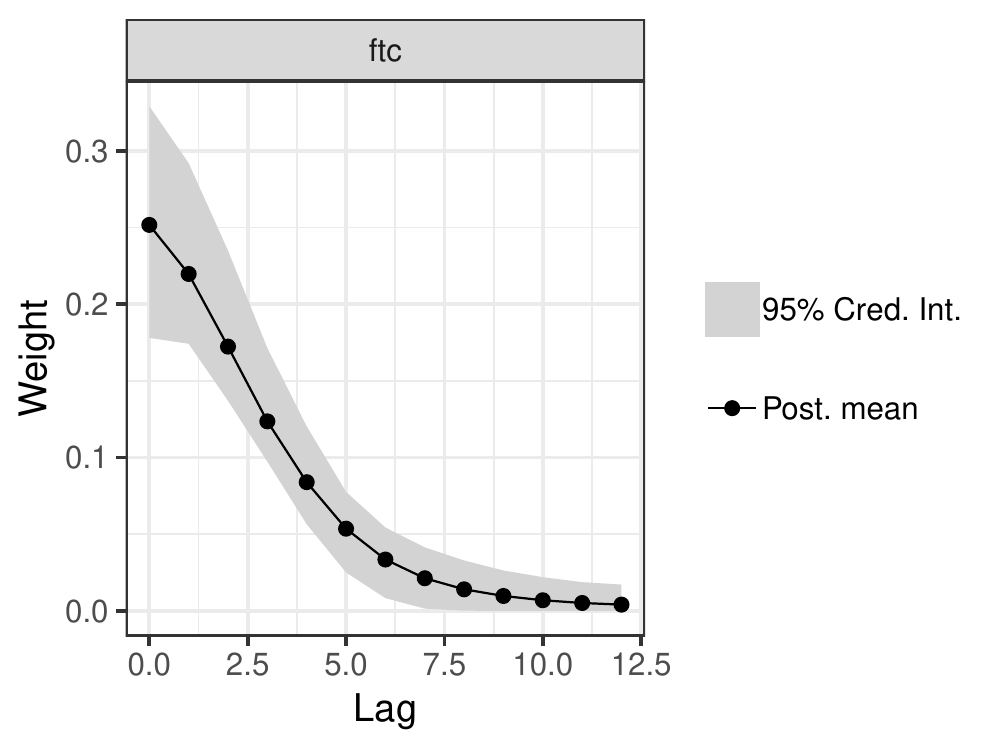}
\caption{Ecological memory of boreal tree growth to forest tent caterpillar defoliation events generated using the \texttt{plotmem()} function.}
\label{fig:ftcmem}
\end{figure}

\begin{figure}
\centering
\includegraphics[width=0.45\textwidth]{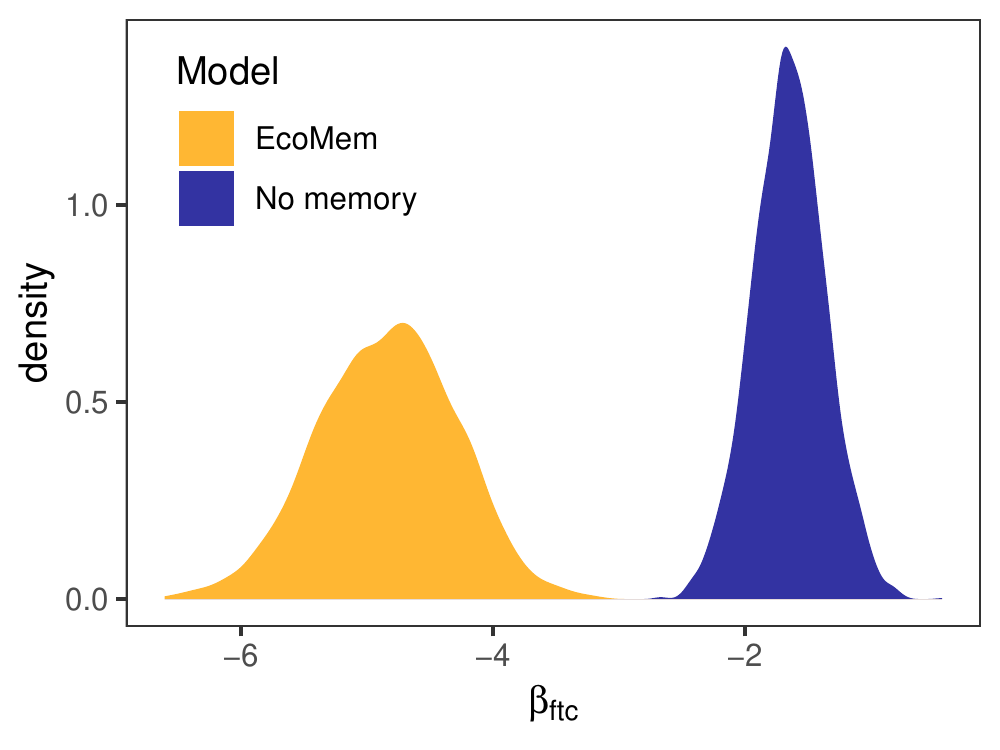}
\caption{Posterior distributions for the effect of forest tent caterpillar defoliation on the mean annual basal area increment of aspen trees with and without accounting for ecological memory.}
\label{fig:ftceffect}
\end{figure}

There is strong evidence of ecological memory in boreal tree growth responses to past forest tent caterpillar defoliation. Posterior mean weight values above 0.01 exist for lags 0 to 8 reflecting persistent impacts of defoliation on mean annual basal area increment (Fig. \ref{fig:ftcmem}). Weighted defoliation event observations were negatively related to the mean annual basal area increment of aspen trees within study sites (Fig. \ref{fig:ftceffect}). The effect of defoliation on basal area increment is much less pronounced if an alternative linear regression model is applied that does not account for ecological memory to defoliation (Fig. \ref{fig:ftceffect}). The stronger effect of defoliation observed after accounting for ecological memory likely reflects the accumulation of defoliation stress on aspen growth over time.

\section{Limitations/Extensions}\label{sec:limits}
High autocorrelation in a memory covariate may lead to poorly identified ecological memory functions. This is particularly evident for covariates that are smooth functions in time with high autocorrelation coefficients for more than 5-10 lags. Future work will focus on modeling weights orthogonal to covariate values reducing potential temporal confounding attributable to high autocorrelation in memory covariates \citep[\textit{sensu}][]{Hanks2015}. The current version of the \ecomem{} package requires users to specify a maximum lag value $L$ for each memory covariate. In some applications, the length of memory may be the main inferential goal. Estimating $L$ within the Bayesian hierarchical model presented in Section \ref{subsec:model} is complicated given realizations of $L$ define a unique set of basis functions used to estimate weights. We are actively working to estimate $L$ as part of \ecomem{}.

\section{Conclusion}\label{sec:conclusion}
Ecological memory and its role in shaping ecosystem responses to global change is an important area of ecological research. We designed the \ecomem{} package to provide ecologists with easily implemented tools to test and account for ecological memory within their respective study systems. We hope that accounting for ecological memory when it is present will lead to improved understanding of the factors contributing to resistent and resilient ecosystem function in the face of changing global conditions. Users are encouraged to apply \ecomem{} widely and report any bugs, issues, or desired extensions on our active issues page (\href{https://github.com/msitter/ecologicalmemory.git}{https://github.com/msitter/ecologicalmemory}).

\begin{table}[h!]
\caption*{\large{\textbf{Availability of software}}}
\begin{tabular}{ll}
\hline
Name of software & \ecomem{}\\
Type of software & Add-on package for R \href{https://cran.r-project.org}{https://cran.r-project.org}\\
First available & 2019\\
Program languages & \texttt{R}\\
Requires & \texttt{R} version 3.5.0 or later including developer tools\\
License & GPL $\geqslant$ 2\\
Code repository & \href{https://github.com/msitter/EcoMem.git}{https://github.com/msitter/EcoMem.git}\\
Installation in \texttt{R} & \verb|devtools::install_github("msitter/EcoMem")|\\
Developer & Malcolm S. Itter, \href{malcolm.itter@helsinki.fi}{malcolm.itter@helsinki.fi}\\
\multirow{2}{*} {Contact address} & Research Centre for Ecological Change, PL 65 (Viikinkaari 1)\\
& 00014 University of Helsinki, Finland\\
\hline
\end{tabular}
\label{tab:soft_avail}
\end{table}

\section*{Acknowledgements}
This work was supported by the Jane and Astos Erkko Foundation and by the National Science Foundation (grant numbers EF-1137309, EF-1241874, EF-1253225).

\section{Authors' contributions}
MI conceived and developed the model framework and wrote the \texttt{R} package. JV and AF assisted with MCMC computational issues. All authors made significant contributions to the manuscript and approved the final draft for submission.

\section*{Supporting Information}

\noindent\textbf{Appendix A} Ecological memory model inference.

\noindent\textbf{Appendix B} Annotated code for simulated and applied case study examples.

\bibliographystyle{apalike}
\bibliography{EcoMem}

\end{document}